\documentclass[aps,prl,psfrag,twocolumn,showpacs,preprintnumbers,amsmath,amssymb]{revtex4}
\usepackage{graphicx}
\usepackage{psfrag}
\usepackage{color} 
\usepackage{dcolumn}
\usepackage{bm}

\newcommand{\tr}{\ensuremath{{\rm tr}}}
\newcommand{\e}{\ensuremath{\text{e}}}

\def\braket#1{\mathinner{\langle{#1}\rangle}}

\begin{document}

\preprint{}

\title{Mesoscopic Charge Relaxation}

\author{Simon~E.~Nigg$^1$, Rosa L\'opez$^2$ and Markus~B\"uttiker$^1$}
\affiliation{$^1$D\'epartement de Physique Th\'eorique, Universit\'e de
  Gen\`eve, CH-1211 Gen\`eve 4, Switzerland}
\affiliation{$^2$Departament de F{\'i}sica,
  Universitat de les Illes Balears, E-07122 Palma de Mallorca, Spain}
\date{\today}

\begin{abstract}
We consider charge relaxation in the mesoscopic equivalent of an RC circuit. 
For a single-channel, spin-polarized contact, self-consistent
scattering theory predicts a universal charge relaxation resistance
equal to half a resistance quantum independent of the transmission
properties of the contact. This prediction is in good agreement with
recent experimental results. We use a tunneling Hamiltonian formalism
and show in Hartree-Fock approximation, that at zero temperature the
charge relaxation resistance is universal even in the presence of
Coulomb blockade effects. We explore departures from universality as a function of temperature and magnetic field.
\end{abstract}

\pacs{85.35.Gv, 73.23.Hk, 73.21.La, 73.22.Dj}

\maketitle
\begin{figure}[b]
\includegraphics[width=0.45\textwidth]{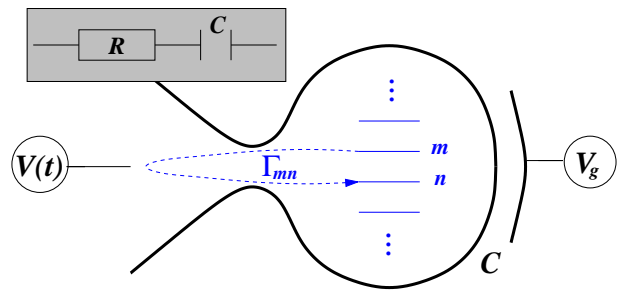}
\caption{\label{fig:dot} Schematics of the mesoscopic capacitor. A
  cavity is connected via one lead to an electron reservoir at voltage
  $V(t)$ and capacitively coupled to a backgate with voltage
  $V_g$. The coupling matrix elements $\Gamma_{mn}$ are
  defined in the text. The
  inset shows the corresponding classical RC circuit.}
\end{figure}

There is increasing interest in the dynamics of 
mesoscopic structures motivated by the desire to manipulate and measure quantum phenomena as rapidly as possible.  It is thus of great importance to characterize the time scales governing the
electron dynamics in simple mesoscopic structures. An elementary but
fundamental building block is the quantum coherent capacitor~\cite{Buettiker:93a}. As in the
classical case, the low frequency dynamics of a mesoscopic capacitor is determined  by a charge relaxation time $\tau_{RC}$. For a quantum coherent capacitor the RC time can still be written as the product of a
resistance and a capacitance, i.e. $\tau_{RC}=R_qC_{\mu}$.
However, due to the coherent nature of electron transport through
mesoscopic structures, both the electrochemical capacitance $C_{\mu}$, which
determines the charge on the capacitor and the charge relaxation
resistance $R_q$, which governs the charge fluctuations, now crucially depend on coherence properties of the system. The capacitance $C_{\mu}$
is related to the imaginary part of the AC conductance but it can also
be obtained by the differentiation of a thermodynamic
(grand-canonical) potential
\cite{Flensberg:93a,Matveev:95a,Buettiker:96b,Aleiner:02a,DutySillanpaa:05a}. The charge
relaxation resistance $R_q$ is related to the real part of
the AC conductance and therefore requires a dynamic theory. In analogy to the classical RC circuit depicted in the upper left corner of
Fig.~\ref{fig:dot}, one has for the mesoscopic system
\begin{equation}\label{eq:1}
G(\omega) = -i\omega C_{\mu}+\omega^2 C_{\mu}^2R_q+O(\omega^3).
\end{equation}
This equation will be taken as a definition of $C_{\mu}$
and $R_q$. If the cavity-reservoir connection permits the transmission of only a single spin polarized channel, a self-consistent
scattering matrix approach, gives at
zero temperature a resistance equal to half a resistance quantum~\cite{Buettiker:93a}
\begin{equation}\label{eq:2}
R_q = \frac{h}{2e^2}.
\end{equation}
We emphasize that the factor of 2 is not connected to spin but 
is rather due to the fact that the cavity connects to only one electron reservoir. More astonishing, even counter-intuitive, is the fact that Eq.~(2) is independent of the transmission properties of the 
channel.  

In a seminal experiment, J.~Gabelli et al.~\cite{Gabelli} have
recently measured both the in
and out of phase parts of the AC conductance of a mesoscopic
RC circuit. In their experiment one ``plate'' of the capacitor consists of a sub-micrometer quantum
dot (QD) and the other is formed by a macroscopic top gate. The role
of the resistor is played by a tunable quantum point contact (QPC)
connecting the QD to an electron reservoir. The results of this experiment
are in good agreement with the theoretical predictions of~\cite{Buettiker:93a}, in particular they confirm 
the universality of the single channel charge relaxation resistance by
using a strong magnetic field to spin polarize the electrons. 
However, it is a priori unclear whether the results derived in
Ref.~\cite{Buettiker:93a} still hold in the presence of single charge effects~\cite{Flensberg:93a,Matveev:95a,Aleiner:02a} which must become important if the 
transmission through the QPC becomes small. Indeed the experiment
observes Coulomb blockade oscillations of the capacitance 
as a function of the gate voltage. It is the aim of the present work
to present a theoretical description for the charge relaxation resistance in the
presence of Coulomb blockade effects. 

The mesoscopic RC circuit along with
the principal model parameters is shown in Fig.~\ref{fig:dot}. $V(t)=V_{\text{ac}}\cos(\omega t)$ is the time
dependent voltage applied to the electron reservoir while $V_g$ is the
voltage applied to the gate and $C$ is the geometrical capacitance
between the QD and the gate. The matrix elements $\Gamma_{mn}$, to be defined below,
describe the indirect coupling between the dot states $m$ and $n$ via
the reservoir.

Our first goal is to show that in the single channel case $R_q$ is universal also in the Coulomb 
blockade regime. For that purpose we treat the cavity at the Hartree-Fock (HF) level~\cite{Brouwer:05a, Hackenbroich:01a}. The starting point of our calculation is the relation
\begin{equation}\label{eq:17}
I(t) = -e\frac{\partial}{\partial t}\Im\{\tr[G^<(t,t)]\},
\end{equation}
which expresses the tunneling current through the QPC as a function of
the lesser (matrix) Green function (GF) $G^<(t,t')$ of the dot. We are interested in the regime where the modulation energy is small 
compared to the level spacing in the dot. Treating the dot as zero
dimensional, the Hamiltonian of this system is~\cite{Platero:03}
\begin{equation}\label{eq:3}
H= H_L + H_D + \sum_{k\sigma,n}(t^{\sigma}_{k,n}c^{\dagger}_{k\sigma}d_{n\sigma} + h.c.),
\end{equation}
where
$H_L=\sum_{k\sigma}E_{k\sigma}(t)c_{k\sigma}^{\dagger}c_{k\sigma}$,
with $E_{k\sigma}(t)=E_{k\sigma}^0+eV(t)$,
describes the (non-interacting) electrons in the isolated lead and $t^{\sigma}_{k,n}$ is the tunneling matrix element
between a reservoir momentum state $k$ and the $n$-th single particle dot state, both with spin
$\sigma$. A change in the gate
voltage is modeled as a shift of the Fermi energy $E_F$ in the reservoir and
we set $V_g=0$. The Hamiltonian of the dot reads
$H_D
=\sum_{n\sigma}\epsilon_{n\sigma}d^{\dagger}_{n\sigma}d_{n\sigma}+E_c(\hat
N_{dot}+\mathcal
N(t))^2$.
Here $E_c=e^2/2C$ is the electrostatic charging energy, $\hat N_{dot}=\sum_{m\sigma}d_{m\sigma}^{\dagger}d_{m\sigma}$ is the
particle number in the dot and $e\mathcal N(t)=C U(t)$ gives
the polarization charges between dot and gate produced by the time
dependent voltage at the reservoir~\cite{Brune:97a}. This polarization
charge in turn, leads to a time-dependent (Hartree) potential $U(t)$
inside the dot and we may write $H_D=\sum_{n\sigma}\tilde\epsilon_{n\sigma}(t)d^{\dagger}_{n\sigma}d_{n\sigma}+E_c{\hat
N_{dot}}^2$, with
$\tilde\epsilon_{m\sigma}(t)=\epsilon_{m\sigma}+eU(t)$. In HF approximation, the
retarded (advanced) GF of the dot takes the form~\cite{Jauho:94,Rosa:01a}
\begin{equation}
G^{R(A)}(t,t') = \e^{i\phi_U(t,t')}G_{eq}^{R(A)}(t-t'),
\end{equation}
where $G_{eq}^{R(A)}(t-t')=G^{R(A)}(t,t')\rvert_{V_{\text{ac}}=0}$ is
the equilibrium retarded (advanced) HF Green function and $\phi_U(t,t') =
\int_t^{t'}d\tau U(\tau)$. The equal time lesser GF is obtained via the Keldysh
equation~\cite{Jauho:94,Wang:99a}
\begin{equation}\label{eq:15}
G^<(t,t) = \int dt_1\int dt_2 G^R(t,t_1)\Gamma^<(t_1,t_2)G^A(t_2,t).
\end{equation}
Here $\Gamma^<(t,t')=i\Gamma\e^{i\phi_V(t,t')}\hat f(t-t')$ with
$\phi_V(t,t') = \int_t^{t'}d\tau V(\tau)$ is the lesser coupling
self-energy and $\hat f(t-t')$ $=$
$(1/2\pi)\int dE\e^{-iE(t-t')}f(E)$, is the Fourier transform of
the Fermi function. $\Gamma^{\sigma}_{mn} = 2\pi\rho_L
{t^{\sigma }_{m}}^*t^{\sigma}_{n}$ are the coupling matrix elements in the wide band
limit~\cite{Jauho:94}. Here and in the following, we use the matrix notation $A_{m\sigma,n\sigma'}\equiv
A^{\sigma}_{mn}\delta_{\sigma\sigma'}$, which takes advantage of the
fact that spin is conserved in (\ref{eq:3}). An important property of the coupling
matrix elements is that
$\Gamma^{\sigma}_{mn}\Gamma^{\sigma}_{kl}=\Gamma^{\sigma}_{ml}\Gamma^{\sigma}_{kn}$,
from which it immediately follows that for  arbitrary matrices $A$ and
$B$
\begin{equation}\label{eq:7}
\tr[\Gamma^{\sigma} A\Gamma^{\sigma}
B]=\tr[\Gamma^{\sigma} A]\tr[\Gamma^{\sigma} B],
\end{equation}
where the trace is over a basis of dot states with spin $\sigma$. Since we are
interested in the linear conductance, we expand (\ref{eq:15}) to
linear order in $V$ and $U$ and find after double Fourier transformation
\begin{equation}\label{eq:16}
G^<(E,E') = G_{eq}^R(E)\Gamma_1^<(E,E')G_{eq}^A(E') + O(U^2,V^2),
\end{equation}
where $\Gamma_1^<(E,E')$ is the double Fourier transform of
$\Gamma_1^<(t,t') = i\Gamma(1+i\phi(t,t'))\hat f(t-t')$ with
$\phi(t,t') \equiv \phi_V(t,t')-\phi_U(t,t')$. With (\ref{eq:17}) and
(\ref{eq:16}) the linear response tunneling current at frequency
$\omega$ becomes $I(\omega) = g(\omega) [V(\omega)-U(\omega)]$, with
\begin{equation}\label{eq:5}
g(\omega)=-i\frac{e\omega}{2\pi}\int dE F(E,\omega)
  \tr[G_{eq}^R(E)\Gamma G_{eq}^A(E-\omega)],
\end{equation}
where $F(E,\omega) = [f(E+\omega)-f(E)]/\omega$ and we have
set $\hbar=1$. To
obtain the AC conductance $G(\omega) = I(\omega)/V(\omega)$, we need
the internal potential $U(\omega)$. For this we note that
in the present single lead system, the displacement current
$-i\omega e\mathcal N(\omega)$ is equal to the tunneling current so that
$g(\omega)[V(\omega)-U(\omega)] =-i\omega C U(\omega)$ and consequently~\cite{Buettiker:96a} $U(\omega) = g(\omega)V(\omega)/[-i\omega
  C+g(\omega)]$. Expanding the conductance to second
    order in frequency, we then obtain after restoring the units
\begin{equation}\label{eq:6}
R_q=-\frac{h}{2e^2}\frac{\int dE\,f'(E)\tr[D(E)^2]}{\left(\int dE\, f'(E)\tr[D(E)]\right)^2},
\end{equation}
where $f'= df/dE$ and  $D(E)\equiv G_{eq}^R(E)\Gamma G_{eq}^A(E)$.
Using (\ref{eq:7}), we have $\tr[D^{\sigma}(E)^2]=\tr[D^{\sigma}(E)]^2$ and hence
at zero temperature
\begin{equation}\label{eq:4}
R_q = \frac{h}{2e^2}\frac{\sum_{\sigma}\nu_{\sigma}(E_F)^2}{\left(\sum_{\sigma}\nu_{\sigma}(E_F)\right)^2},
\end{equation}
where $\nu_{\sigma}(E)\equiv\tr[D^{\sigma}(E)]/2\pi$ is the density of spin
$\sigma$ states in the dot. Eqs.~(\ref{eq:5},\ref{eq:4}) are central
results of this work. In particular, Eq.~(\ref{eq:4}) demonstrates
that for a single (spin polarized) channel $R_q$ is still given by Eq.~(\ref{eq:2}).  

In the following, we use Eq.~(\ref{eq:4}) to investigate the magnetic
field dependence of $R_q$. We consider here a dot with two spin degenerate
levels with bare energies $\epsilon_{1\sigma}$ and
$\epsilon_{2\sigma}=\epsilon_{1\sigma}+\Delta$ respectively. In the
numerical calculations, we set $E_c/\Delta=2.5$. We are interested in the regime of low magnetic field, where
the Zeeman splitting $\Delta_B=\mu_BgB\leq\Delta$. For simplicity, we further assume that $\Gamma^{\uparrow}_{mn}=\Gamma^{\downarrow}_{mn}\equiv\gamma$, for
$m,n\in\{1,2\}$. The equilibrium HF retarded GF of the dot may be
written as $G_{eq}^R(E) = [G_0^R(E)^{-1}-\Sigma_{HF}^{R}+i\Gamma/2]^{-1}$,
with the non-interacting equilibrium GF of the isolated dot
$({G_0^R(E)})_{mn}=\delta_{mn}(E-\epsilon_{m}+i0^+)^{-1}$ and the HF self-energy
\begin{equation}\label{eq:9}
(\Sigma_{HF}^{R})^{\sigma}_{mn}=E_c\Big[\delta_{mn}{\sum_{l\sigma'}}\braket{n_{l\sigma'}}-\braket{d_{m\sigma}^{\dagger}d_{n\sigma}}\Big].
\end{equation}
The most important feature of this self-energy is that it correctly
excludes the unphysical self-interaction terms ($m=n=l$ and
$\sigma=\sigma'$) of the Hartree
approximation and consequently leads to the appearance of the Coulomb
gap across $E_F$, which is the essential spectral signature
of the Coulomb blockade effect. The ``mean fields''
$\braket{d^{\dagger}_{m\sigma}d_{n\sigma}}$ are determined
self-consistently~\cite{Golosov:06} by solving the set of equations
\begin{equation}\label{eq:10}
\braket{d^{\dagger}_{m\sigma} d_{n\sigma}}=\frac{1}{2\pi}\int_{-\infty}^{\infty}dE
f(E)[G_{eq}^R(E)\Gamma G_{eq}^A(E)]^{\sigma}_{mn}.
\end{equation}
Because of the interaction,
$\nu_{\sigma}(E)$ depends on the level occupancies and we must
distinguish two cases.
\begin{figure}
\centering
\includegraphics[width=0.5\textwidth]{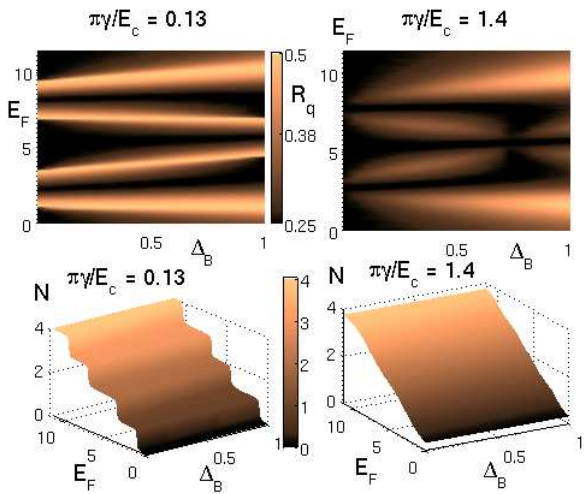}
\caption{\label{fig:Rqspin} Magnetic field dependence of the charge
  relaxation resistance $R_q$. The
  upper panels show $R_q$ as a function of the Zeeman splitting
  $\Delta_B$ and the Fermi energy $E_F$ for weak $\pi\gamma/E_c=0.13$
  and strong $\pi\gamma/E_c=1.4$ coupling. All energies are given in
  units of the bare level spacing $\Delta$. The lower panels show the
  corresponding total dot charge.}
\end{figure}
Solving Eq.~(\ref{eq:10}) numerically in the strongly blockaded regime $\pi\gamma/E_c\ll 1$, we find that
$R_q$ is non-universal even as $B\rightarrow 0$ (upper left panel
of Fig.~\ref{fig:Rqspin}). This is due to the
fact that the dot charge is strongly quantized in this
regime, as shown in the lower left panel of Fig.~\ref{fig:Rqspin}, which leads to a gap of order $E_c+\Delta\delta_{\sigma\sigma'}+(1-\delta_{\sigma\sigma'})\Delta_B$ between the highest occupied
state with spin $\sigma$ and the lowest unoccupied state with spin
$\sigma'$. There are thus four well separated (separation $\sim E_c$),
narrow (width $\sim\gamma$) resonances in the total
density of states $\sum_{\sigma}\nu_{\sigma}$ as a function of
$E_F$. We can understand the particular behavior
of $R_q$ shown in the upper left panel of Fig.~\ref{fig:Rqspin} at
specific values of the Fermi energy. If $E_F$ is on a resonance with spin $\sigma$, then
$\nu_{\sigma'}(E_F)\cong\nu_{\sigma}(E_F)\delta_{\sigma\sigma'}$. Therefore
on resonance, i.e. for $E_F\in\{\epsilon_{1\sigma},\epsilon_{2\sigma},\epsilon_{1\overline{\sigma}}+E_c+\Delta_B,\epsilon_{2\overline{\sigma}}+E_c+\Delta_B\}$,
we expect to have $R_q\cong h/2e^2$ according to
(\ref{eq:4}). Furthermore, in the middle
between two consecutive resonances
$\nu_{\sigma}(E_F)=\nu_{\overline{\sigma}}(E_F)$ and so $R_q$ takes
on its minimal two channel value $h/4e^2$. In the
opposite limit of strong coupling $1\lesssim\pi\gamma/E_c$, the Coulomb blockade is smeared out
and the charge on the dot is not strongly
quantized anymore (see lower right panel in Fig.~\ref{fig:Rqspin}). Thus at very low magnetic fields, the two spin states
are nearly simultaneously charged and the degeneracy is not
appreciably lifted. Therefore,
$R_q\cong h/4e^2$ at low magnetic fields as shown in the upper
right panel of Fig.~\ref{fig:Rqspin}. As $\Delta_B$ increases the broad
resonances in the density of states corresponding to the two spin
degenerate levels split into consecutively overlapping resonances and
$R_q$ starts to increases except in the middle between two
neighboring resonances. Finally, a crossing of the two innermost resonances is
observed for $\Delta_B\cong 0.75$. Such a crossing occurs when the Zeeman splitting approaches the effective level
spacing
$\Delta+E_c(\braket{n_{1\sigma}}-\braket{n_{2\overline\sigma}})$,
which interestingly is seen to be smaller than the bare level
spacing here. This effect is peculiar to the strongly coupled regime and is
due to enhanced exchange interactions, which favor the consequent
population of states with equal spins~\cite{Kurland:01,Rokhinson:01}. At the
crossing point the densities of both spin states are equal and $R_q$ takes on its minimal value.

\begin{figure}
  \psfrag{g=0.1, U=2.5, D=1}[b][]{$\gamma=0.1, E_c=2.5, \Delta=1$}
  \psfrag{Ef}[t][]{$E_F$}
  \psfrag{Rq}[][r][1][-90]{\begin{tabular}{c}$R_q$\\$[\frac{h}{e^2}]$\end{tabular}}
  \psfrag{Rq2}[][][0.8][-90]{$R_q$}
  \psfrag{1/beta}[][][0.8]{$1/\beta$}
  \psfrag{beta=0.02}[][][0.8]{$\quad \beta=0.02$}
  \psfrag{beta=0.5}[][][0.8]{$\quad \beta=0.5$}
  \psfrag{beta=12.5}[][][0.8]{$\quad \beta=12.5$}
  \psfrag{beta=100}[][][0.8]{$\quad \beta=100$}
  \psfrag{e1}[t][][1.2]{$\epsilon_1$}
  \psfrag{e2}[t][][1.2]{$\epsilon_2^*$}
\includegraphics[width=0.45\textwidth]{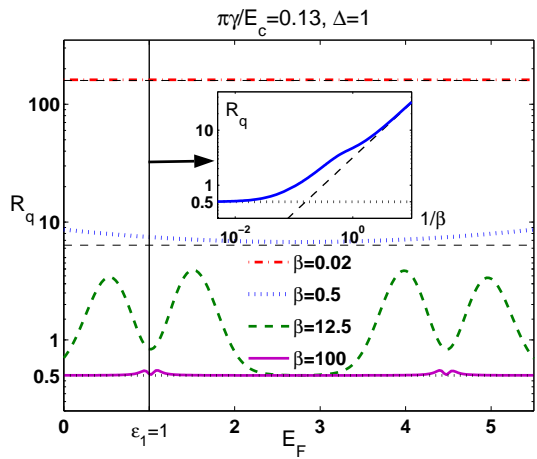}
\caption{\label{fig:RqT} Charge relaxation resistance $R_q$ as a function of Fermi energy $E_F$ for 
different temperatures. The lower three curves are for a two level dot. 
The upper most curve is the high temperature asymptote. The inset shows
$R_q$ as a funtion of temperature for $E_F  = \epsilon_1$. }
\end{figure}

As a second application, we investigate the temperature
dependence of $R_q$ in the high magnetic field limit, where the
incoming electrons are effectively spin polarized and there is only a
single transmitting channel through the QPC. We consider here a
quantum dot with two (bare) levels $\epsilon_1$ and
$\epsilon_2=\epsilon_1+\Delta$ and suppress the now
superfluous spin index. In the low temperature regime $k_BT\ll\gamma$, we may expand $\nu(E)$
and $\nu(E)^2$ around $E_F$ assuming that these functions vary slowly
in the range $E_F\pm k_BT/2$. This yields, to first non-vanishing order in $k_BT\equiv\beta^{-1}$
\begin{equation}\label{eq:12}
R_q \cong
\frac{h}{2e^2}\left(1+\frac{\pi^2}{3\beta^2}\left[\frac{\nu'(E_F)}{\nu(E_F)}\right]^2\right),
\end{equation}
where $\nu'(E)\equiv\frac{\partial\nu(E)}{\partial E}$. Thus in
HF approximation, the lowest order correction is proportional to the square of the energy derivative of the density of states at the
Fermi energy. This explains the presence of the peaks seen in
Fig.~\ref{fig:RqT}, to the left
and right of the two resonances at $E_F=\epsilon_1=1$ and
$E_F=\epsilon_2^*=\epsilon_1+E_c+\Delta$, for the two lowest
temperatures $\beta=100$ and $\beta=12.5$. The fact that for
$\beta=12.5$, $R_q$ does not identically vanish at resonance where $\nu'(E_F)=0$, is due to higher
order terms in the low temperature expansion, which involve
non-vanishing higher order derivatives of $\nu$. The dotted
horizontal line at $R_q=h/2e^2$ marks the zero temperature
result. In the opposite
limit of very high temperature $k_BT\gg\Delta+E_c$, where
$-f'(E)\cong\beta/4$ and in the weakly coupled regime $\gamma\ll \Delta$, where the
density of states is well approximated by a sum of displaced
Lorentzians of width $\gamma$,
we can estimate the remaining integrals in the numerator and
denominator of (\ref{eq:6}) as $-\int dE f'(E)\nu(E)^2\cong \beta
  M/4\pi\gamma$ and $-\int dE f'(E)\nu(E)\cong
\beta M/4$ respectively, where $M$ is the number of density of states peaks under the
broad curve $f'(E)$. Therefore at very high
temperature we find
\begin{equation}\label{eq:13}
R_q\cong\frac{h}{e^2}\frac{2}{\pi\gamma M\beta}.
\end{equation}
This high temperature limit is shown as a dashed horizontal black line in
Fig.~\ref{fig:RqT} for the two highest temperatures. In the present
two level system, $M=2$ and as
expected the agreement with the numerical integration of (\ref{eq:6})
is good for the highest temperature
$\beta^{-1}=50\gg\Delta+E_c=3.5$. For the intermediate temperature
$\Delta=1<\beta^{-1}=2<\Delta+E_c=3.5$, there is already a significant
deviation from the asymptotic result. In the inset of
Fig.~\ref{fig:RqT}, we show $R_q$ as a function of the temperature on the first resonance at $E_F=\epsilon_1$. The dashed line
corresponds to the high temperature asymptote \eqref{eq:13} and the dotted line
is the low temperature limit~\eqref{eq:12}.

In this work we have analyzed the charge relaxation of a mesoscopic
capacitor in the linear regime of coherent dynamical transport. We have shown that the single
channel zero temperature charge relaxation resistance $R_q$ is
universal even in the presence of single charge effects, described in
the Hartree-Fock approximation. This shows in particular that charge relaxation of a
quantum coherent capacitor is faster than one could naively expect
based on classical arguments. We obtain the magnetic
field dependence of $R_q$ in the two channel case (electrons with spin),
where we identify two qualitatively different regimes of weak and
strong coupling. In the former, the degeneracy of both spin states is
lifted by the interaction at all field strengths and $R_q$ is non-universal. In
the latter regime, the degeneracy is lifted only at finite field and
at zero field $R_q$ is universal and equal to its minimum two channel
value $h/4e^2$. The finite temperature behavior of $R_q$
for a two level spin polarized system is, to lowest order,
determined by the logarithmic derivative of the density of states with respect to energy. In the
multilevel case the HF approximation gives a
reasonable qualitative picture of the underlying physics. The
important case of $B=0$, for a single strongly coupled level in the
dot, requires a treatment of Kondo physics and 
will be discussed elsewhere~\cite{unpub}.

Our work demonstrates that mesoscopic charge relaxation is a physically very
interesting process and provides a basis for the understanding of
experimental data in the low and high magnetic field ranges. 

We thank C. Glattli and J. Gabelli for discussions and sharing their data with us.
This work was supported by the Swiss NSF, MaNEP, the Spanish MEC by
MAT2005-07369-C03-03 and the Ramon y Cajal program. 


\end{document}